# Promises and technological prospects of two-dimensional Rashba materials

Arjyama Bordoloi,[1] A. C. Garcia-Castro,[2] Zachary Romestan,[3] Aldo H. Romero,[3] and Sobhit Singh[1, 4, *]

[1]*Department of Mechanical Engineering, University of Rochester, Rochester, New York 14627, USA*
[2]*School of Physics, Universidad Industrial de Santander,
Carrera 27 Calle 9, 680002 Bucaramanga, Colombia*
[3]*Department of Physics and Astronomy, West Virginia University, Morgantown, WV 26505-6315, USA*
[4]*Materials Science Program, University of Rochester, Rochester, New York 14627, USA*
(Dated: April 20, 2024)

The Rashba spin-orbit coupling effect, primarily arising from structural-inversion asymmetry in periodic crystals, has garnered considerable attention due to its tunability and potential applications in spintronics. Its capability to manipulate electron spin without an external magnetic field opens new avenues for spintronic device design, particularly in semiconductor technology. Within this framework, 2D Rashba materials hold special interest due to their inherent characteristics, which facilitate miniaturization and engineering capabilities. In this Perspective article, we provide an overview of recent advancements in the research of 2D Rashba materials, aiming to offer a comprehensive understanding of the diverse manifestations and multifaceted implications of the Rashba effect in material science. Rather than merely presenting a list of materials, our approach involves synthesizing various viewpoints, assessing current trends, and addressing challenges within the field. Our objective is to bridge the gap between fundamental research and practical applications by correlating each material with the necessary advancements required to translate theoretical concepts into tangible technologies. Furthermore, we highlight promising avenues for future research and development, drawing from insights gleaned from the current state of the field.

## I. INTRODUCTION

Spintronics is a rapidly burgeoning research field with the potential to supersede conventional electronics in seamlessly integrating memory, processing, communication, and sensing by utilizing electrons' spin degree of freedom as a logical unit [1–9]. The field of spintronics represents a significant shift in the approach to electronic devices, focusing on the transport of electron spin rather than on the conventional flow of charge current. Various methods, including thermal, optical, electric, and magnetic techniques, have facilitated this paradigm shift, contributing to the versatility and potential of spintronic applications [10, 11]. However, the design of most spintronic devices relies heavily on the principles of giant magnetoresistance (GMR) [12, 13] and tunneling magnetoresistance (TMR) [14]. These principles, first identified in the late 1980s and the early 1990s, have been foundational to developing spintronics [15–20]. GMR involves significant changes in the electrical resistance in response to an external magnetic field in layered ferromagnetic materials. Conversely, TMR is a phenomenon in which the resistance of a magnetic tunnel junction changes depending on the relative alignment of the magnetization in ferromagnetic layers separated by an insulating barrier.

A common characteristic of most of the spintronic devices is ferromagnetic materials. While ferromagnetic materials are integral to the functioning of GMR- and TMR-based devices, they pose several challenges, particularly regarding device integration [20–24]. Integration issues stem from factors such as the incompatibility of ferromagnetic materials with standard semiconductor processes, difficulties in controlling their magnetic properties at the nanoscale, and challenges in maintaining consistent performance under varying operational conditions. In response to these challenges, semiconductor spintronics has emerged as a promising alternative. This approach leverages the properties of semiconductors combined with spin-orbit coupling (SOC) to generate and manipulate spin currents. SOC is a relativistic effect that arises from the interaction between an electron's spin and its orbital motion.

Among the various SOC effects explored in semiconductor spintronics, Rashba spin-orbit coupling (RSOC) has gained considerable attention [25–28]. RSOC is a phenomenon in which an external electric field can tune the SOC-induced effects, making it highly adaptable for spintronic device applications. This tunability is a crucial advantage of RSOC over other SOC effects because it allows the manipulation of spin currents without requiring an external magnetic field. This independence from magnetic fields is particularly beneficial in semiconductor spintronics because it enables the design of spintronic devices using nonmagnetic materials. This approach circumvents the integration challenges of ferromagnetic materials and opens new avenues for device functionality and miniaturization [29–32].

The concept of RSOC was initially proposed by Emmanuel Rashba in 1959 [25]. However, it gained significant prominence in the scientific community following two publications in 1984 by Bychkov and Rashba [27, 28] in response to the experimental findings related to the Quantum Hall Effect (QHE) [33, 34]. These foundational papers, published in the early 1980s, provide crucial insights into the Quantum Hall Effect, a phenomenon observed in two-dimensional (2D) electron systems under strong magnetic fields, which later played a pivotal role in understanding various quantum phenomena.

* s.singh@rochester.edu

Originally, RSOC was believed to be exclusive to 2D surfaces and interfaces. This assumption was based on the understanding that RSOC arises from structural inversion asymmetry (SIA) in these confined systems [21]. The relativistic motion of electrons in an asymmetric crystal potential induced by the SIA leads to the generation of an effective magnetic field, denoted as $\mathbf{B}_{\text{eff}}$. This field acts analogously to an external magnetic field but is intrinsic to material's crystal structure. However, subsequent research has significantly expanded our understanding of RSOC, revealing its presence even in bulk three-dimensional crystal structures [28, 35–39], and in centrosymmetric crystals (hidden Rashba effect) [40–45]. This revelation, supported by various studies conducted in the $21^{st}$ century, underscores the ubiquity and versatility of RSOC in different material systems [46].

The effective magnetic field $\mathbf{B}_{\text{eff}}$ resulting from the SIA has profound implications for the behavior of electron spins in nonmagnetic materials. When electrons are subjected to this field, their spins experience torque, causing them to precess. This precession is the physical manifestation of the RSOC and is a critical factor in manipulating spin currents in spintronic devices [22].

Spin field-effect transistors (s-FETs), first proposed by Datta and Das in 1990, mark one of the significant advancements towards applying RSOC in practical spintronic devices [47]. In s-FETs, spins injected from a ferromagnetic (FM) source are transported to the ferromagnetic drain by their precessional motion around the magnetic field governed by the RSOC in two-dimensional electron gas (2DEG) hosted by a semiconducting channel, as schematically shown in Figure 1. The spin-precession length and, hence, the spin orientation of the carriers can be modulated by an external gate voltage that tunes the Rashba spin-splitting parameter.

Apart from this, the interplay of RSOC with superconductivity brings up the possibility of realizing various exotic quantum states, which might have a revolutionary impact in the field of quantum information processing [48–50]. The combination of 2D Rashba materials with s-wave superconductors under broken time-reversal symmetry is predicted to host topologically-protected Majorana edge modes [51, 52]. These Majorana states are unique and robust against certain types of perturbations, making them promising candidates for fault-tolerant topological quantum computation if realized experimentally [53–55].

In the context of spintronic device applications, 2D Rashba materials are particularly attractive due to their inherent characteristics, facilitating miniaturization and engineering capabilities. In this Perspective article, we aim to offer an overview of recent progress in researching 2D Rashba materials. Starting with Section II, which delves into the fundamental theoretical concepts essential for understanding the Rashba effect, Section III provides a comprehensive summary of the recent advancements in the study of various 2D Rashba materials. These include AB binary monolayers, transition metal dichalcogenides (TMDs), Janus TMDs, and other Janus monolayers, as well as van der Waals (vdW) heterostructures. Additionally, Section III explores diverse strategies for manipulating Rashba spin splitting in materials. This includes applying external electric fields, strain engineering, and the influence of substrate proximity – critical factors in identifying materials suitable for prospective spintronic device design. Furthermore, the role of RSOC in governing the nontrivial-topological electronic phase transition in specific 2D materials is discussed in Subsection III F. The optical manipulation of RSOC is included in the subsequent Subsection III G. Section IV presents an overview of the practical applications of RSOC in spintronic technology, followed by a Summary and Outlook in Section V.

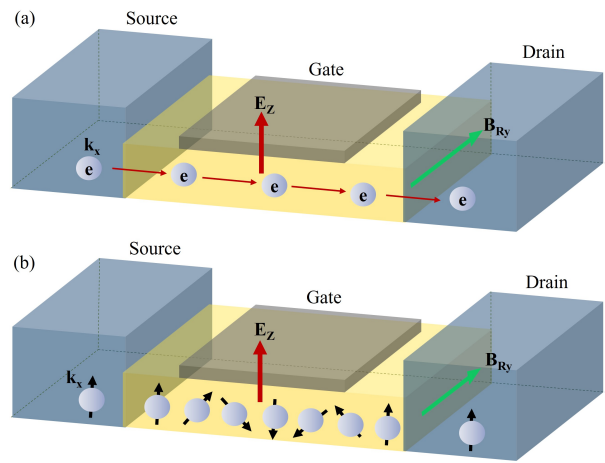

FIG. 1. (a) (Color online) Schematic representation of electrons traversing from a ferromagnetic source to a ferromagnetic drain within 2DEG hosted by a semiconducting channel, in the presence of a perpendicular electric field $E_z$ that induces a Rashba magnetic field $\mathbf{B}_{\mathbf{Ry}}$. (b) Precessional spin movement of electrons around the Rashba magnetic field while traveling from the source to the drain, modulated by the gate voltage.

## II. THEORETICAL FOUNDATION

SOC is a relativistic phenomenon that lifts the spin degeneracy of electronic bands provided both space-inversion (I) and time-reversal (TR) symmetries are not conserved simultaneously [56–59]. In the case of free electron gas, the presence of I and TR symmetries mandates the parabolic nature of electronic bands to retain their spin degeneracy, as illustrated in Fig. 2(a). The TR symmetry dictates $E(\mathbf{k},\uparrow) = E(\mathbf{-k},\downarrow)$, while the I symmetry mandates $E(\mathbf{k},\uparrow) = E(\mathbf{-k},\uparrow)$, where $E$ and $\mathbf{k}$ denote the energy eigenvalues and crystal momentum, respectively. The up and down arrows represent the spin-up and spin-down states. The breaking of inversion symmetry (while preserving TR symmetry) induces spin splitting of energy bands at all generic $\mathbf{k}$ points, except at the time-reversal-invariant momenta (TRIM) points, where Kramer's degeneracy holds [60].

In nonmagnetic polar materials, *i.e.*, preserved TR but broken I symmetry, the RSOC can be described as a linear coupling between electron's spin $\boldsymbol{\sigma}$ and crystal momentum $\mathbf{k}$. A naive concept of RSOC can be illustrated by considering the case of a free 2DEG [29]. A relativistic electron of mass $m$ with momentum $\mathbf{k}$ moving in an electric field ($\mathbf{E} = E\hat{z}$), acting in the direction of broken inversion symmetry experiences a magnetic field $\mathbf{B}_{\text{eff}} = -\frac{\hbar}{mc^2}(\mathbf{k} \times \mathbf{E})$, where $\hbar$ is reduced Planck's constant and $c$ is the speed of light. The Zeeman interaction of electron's spin moment with the effective magnetic field mimics the form of RSOC. Thus, the Rashba Hamiltonian ($\hat{H}_R$) can be expressed as

$$\hat{H}_R = \frac{e\hbar}{2m}(\boldsymbol{\sigma} \cdot \mathbf{B}) = \alpha_R(\boldsymbol{\sigma} \times \mathbf{k}) \cdot \hat{\mathbf{z}}, \quad (1)$$

where $\alpha_R$ is the Rashba parameter, which represents the strength of the SOC, and it is a crucial design parameter on Rashba materials, as it represents the strength of the spin splitting. The Hamiltonian for a 2DEG, including the Rashba term, reads:

$$\hat{H} = \frac{\hbar^2 k^2}{2m} + \alpha_R(\boldsymbol{\sigma} \times \mathbf{k}) \cdot \hat{\mathbf{z}}. \quad (2)$$

Solving equation $\hat{H}|\psi\rangle = E|\psi\rangle$ yields following energy eigenvalues and eigenstates.

$$E_\pm = \frac{\hbar^2 k^2}{2m} \pm \alpha_R k = \frac{\hbar^2}{2m}(k \pm k_R)^2 + E_R, \text{ and} \quad (3)$$

$$|\psi_\pm(k)\rangle = \frac{1}{\sqrt{2}}\begin{pmatrix} \pm e^{-i\theta} \\ 1 \end{pmatrix}. \quad (4)$$

Figure 2(b) represents a prototypical electronic band structure illustrating the lifting of spin degeneracy due to RSOC. The offset Rashba momentum $k_R$ ($= m\alpha_R$) and Rashba energy $E_R$ ($= m\alpha_R^2/2$) can be determined computationally and the Rashba parameter $\alpha_R$ can be calculated using the expression $\alpha_R = 2E_R/k_R$. Furthermore, angle $\theta$ in equation 4 can be represented as $\tan^{-1}(k_y/k_x)$. On the other hand, the experimental observation of RSOC is possible using either electron photoemission or transport experiments.

The spin texture is obtained by calculating the expectation value of the spin operator $\boldsymbol{\sigma}$, expressed as

$$\langle\psi_\pm(k)|\boldsymbol{\sigma}|\psi_\pm(k)\rangle = \pm\begin{pmatrix} \cos\theta \\ -\sin\theta \\ 0 \end{pmatrix}. \quad (5)$$

Hence, the spin texture is independent of the magnitude of $\mathbf{k}$; rather, it depends on its direction within the plane of the 2DEG. As $\mathbf{k} \to -\mathbf{k}$, the angle $\theta$ varies from 0 to $\pi$, reversing the spin orientation. This ensures the conservation of TR symmetry in pure Rashba materials.

Often, materials possessing high-symmetry surfaces, such as those characterized by point groups $C_{3v}$ and $C_{4v}$, exhibit anisotropic behaviour in their Rashba spin splitting, which may not be captured within the conventional Rashba model having linear-momentum dependence. In such cases, the Hamiltonian must incorporate terms up to the third-order in $\mathbf{k}$ [61, 62].

## III. RASHBA EFFECT IN 2D MATERIALS

In this Perspective article, our primary focus is on conventional 2D Rashba materials. Our discussion aims to provide a holistic view of the diversity of Rashba effect and its multifaceted implications in material science. Through this approach, we aim to enrich the discourse on Rashba materials, offering insights into the conventional Rashba materials and thereby contributing to the broader understanding and application of Rashba physics in advanced materials development.

Beyond transition metal dichalcogenides (TMDs) and group-IV monolayers such as silicene, germanene, and stanene (as discussed below in this section), various emerging classes of 2D materials like halide perovskites [68] and topological insulators also showcase Rashba-induced phenomena [30]. These materials, characterized by high carrier mobility, adjustable band gaps, and robust light-matter interactions, are further empowered by the Rashba effect to support innovative device functionalities. For example, the manipulation of spin texture and spin-momentum locking in topological insulators, facilitated by the Rashba effect, enables the development of new quantum devices that leverage the topological protection of edge states for efficient, low-dissipation transport [69–72]. Exploring the Rashba effect in 2D materials extends beyond practical applications to deepen our understanding of fundamental physics, such as the dynamics between spin-orbit coupling and electron-electron interactions in reduced dimensions [73–78]. This foundational knowledge is vital for advancing quantum materials and devices grounded in topological physics and spintronics principles. The Rashba effect in 2D materials expands the functional material library for cutting-edge applications. It pushes the boundaries of our theoretical and experimental grasp of spin-orbit phenomena in condensed matter physics [79].

Below we provide a comprehensive overview of various 2D materials exhibiting Rashba effect. Our discussion is not just a catalog of materials; it represents a synthesis of our viewpoints with the current trends and challenges within the field. We aim to bridge the gap between fundamental research and practical applications by correlating each material with the necessary advancements to transition theoretical concepts into tangible technologies. This analysis encompasses an assessment of the current state of the field, identifying promising avenues for future research and development.

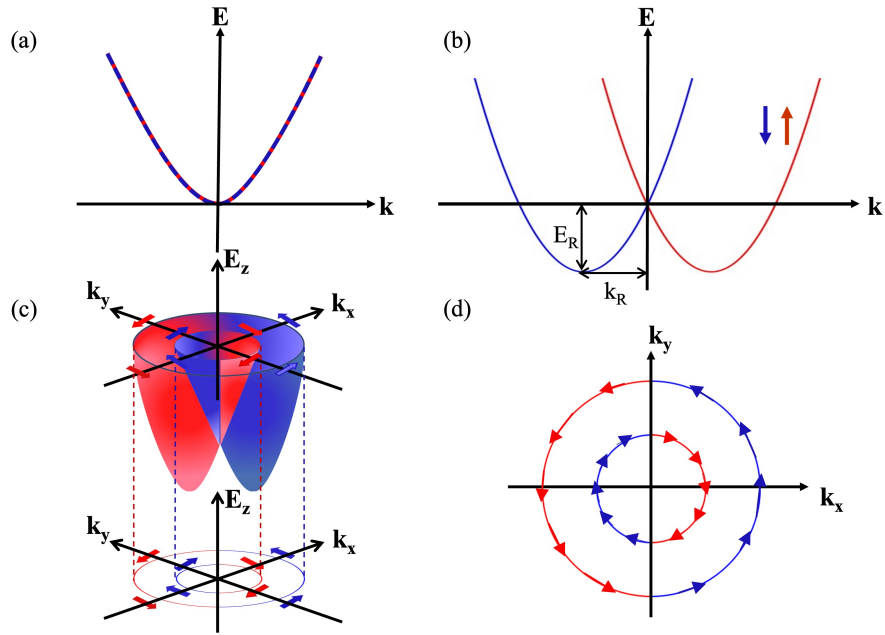

FIG. 2. (a) (Color online) Schematic representation of spin degenerate electronic band structure for 2DEG without SOC. (b) Rashba spin splitting resulting in spin degeneracy at all **k** points, except at the TRIM point. $E_R$ and $k_R$ denote the Rashba energy and Rashba momentum offset, respectively. (c) Schematic 3D representation of electronic band structure in the presence of the Rashba effect. (d) Helical spin texture at a constant energy surface for the case of pure Rashba spin splitting.

### A. $AB$ binary buckled monolayers

Centrosymmetric 2D phosphorenes do not exhibit the Rashba effect. However, RSOC can be induced by alloying phosphorene with heavier elements. The resulting buckled hexagonal P$X$ ($X$ = As, Bi, and Sb) monolayers, possess significant Rashba splitting near the $\Gamma$ point of conduction band minimum (CBM) and $\alpha_R$ shows an increasing trend from 0.13 to 1.56 eV Å with increasing atomic number of $X$ in agreement with the expected increasing of the SOC strength [80]. Moreover, the stable free-standing buckled honeycomb monolayers of BiSb and AlBi exhibit a large and tunable RSOC with $\alpha_R$ of 2.3 [64] and 2.77 eV Å [81], respectively. In the AlBi monolayer, RSOC is sensitive to strain, while in the BiSb monolayer, it is sensitive to both strain and electric field [81]. Wu et al. [81] designed a spin field-effect transistor (s-FET) based on BiSb monolayer and reported a short spin channel length (42 nm, tunable with strain) compared to conventional s-FETs (about 2–5 $\mu$m).

Isostructural monolayers of $h$-NbN and $h$-TaN are reported to exhibit a substantial Rashba spin splitting, with $\alpha_R$ values of 2.9 and 4.23 eV Å, respectively [82]. The $h$-TaN monolayer has a higher value of $\alpha_R$ compared to $h$-NbN due to the larger SOC induced by the heavier Ta atoms [82]. Among the monolayer Mg$X$ (X=S, Se, Te) family, MgTe exhibits the highest Rashba spin splitting with $\alpha_R$ of 0.63 eV Å [83]. Likewise, ZnTe and CdTe exhibit moderate Rashba spin splitting with $\alpha_R$ of 1.06 eV Å and 1.27 eV Å, respectively [84]. Based on the first-principles calculations, Liu et al. [85] report moderate $\alpha_R$ values of 0.60, 0.62, and 0.60 eV Å for GeTe, SnTe, and PbTe monolayers, respectively. Rehman et al. [86], on the other hand, investigated the Rashba properties of $MX$ monolayers ($M$ = Mo, W; and $X$ = C, S, Se) and determined the Rashba parameter near the $\Gamma$ point of valence band maximum (VBM) to be 0.14, 1.02, 1.2, and 1.26 eV Å, for MoC, WC, WS, and WSe monolayers, respectively. Remarkably, these monolayers demonstrated a quantum valley Hall effect due to their distinctive Berry curvatures [86]. Figure 3(a) shows the crystal structure of the binary buckled square monolayers of Pb$X$ (where $X$ = S, Se, and Te), which inherently exhibit RSOC. Specifically, the buckled square PbS monolayers exhibit $\alpha_R$ values of 1.03 and 5.10 eV Å near the $\Gamma$- and $M$-points of Brillouin zone for the conduction band minimum (CBM), respectively. However, the Rashba effect is absent in the planar square PbS monolayer due to the presence of inversion symmetry [63].

Generally, the intrinsic RSOC present in 2D materials can be effectively tuned by applying external perturbative methods, including electric fields and strain. This tunability makes these materials suitable candidates for spintronic devices. PBi monolayer, which exhibits the highest Rashba splitting among all P$X$ monolayers, shows a variation of $\alpha_R$ from 1.56 to 4.41 eV Å under 10 % biaxial tensile strain [80]. In BiSb, $\alpha_R$ increases from 2.3 eV Å to 3.56 eV Å at 6 % biaxial tensile strain and reduces to 1.77 eV Å at 4% compressive strain. Furthermore, when subjected to strain, BiSb monolayer, crystal structure shown in Fig. 3(b), changes from a direct to an indirect bandgap semiconductor [64]. On the contrary, RSOC in GaTe monolayer is nearly insensitive to biaxial strain [87]. The $\alpha_R$ in MgTe monolayer can be tuned up to ±0.2 eV Å under applied biaxial strain [83]. Similarly, applying a positive electric field or compressive strain can significantly enhance the $\alpha_R$ in $MX$ monolayers. Conversely, applying a negative electric field or tensile strain weakens the Rashba effect in these materials [86].

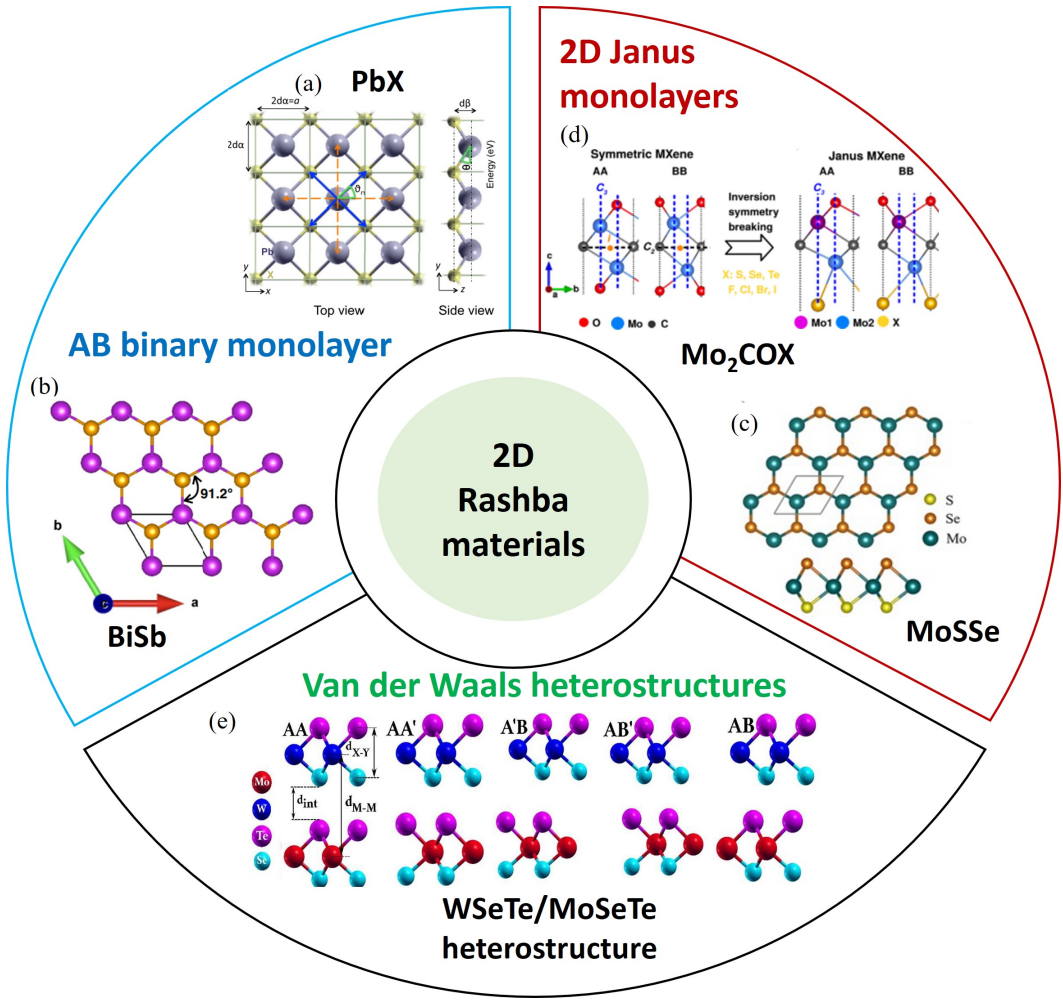

FIG. 3. Some of the typical 2D Rashba materials. (a) Structure of 2D square buckled PbX (X = S, Se, Te) monolayer from top and side views, employing distinct colors for different atom types [63]. (b) 2D hexagonal buckled configuration of BiSb, where purple and pink balls symbolize Bi and Sb atoms, respectively [64]. (c) Structure of Janus MoSSe, a representative structure of the family of 2D Janus TMDs [65]. (d) Left: AA and BB configuration of symmetrically passivated $Mo_2CO_2$ from the side view. Right: Non-centrosymmetric AA and BB terminating configurations of Janus $Mo_2COX$ (X = S, Se, Te; F, Cl, Br, I) [66]. (e)Side view of the Janus bilayer of WSeTe/MoSeTe for five stacking orders [67].

### B. TMDs and Janus TMDs

The hexagonal 2D transition-metal dichalcogenides (TMDs) ($MX_2$, $M$ = Mo, W, and $X$ = S, Se, Te) do not exhibit intrinsic RSOC due to their out-of-plane mirror symmetry. However, this symmetry can be broken and RSOC can be induced in these materials by replacing one of the chalcogen atoms, $X$, with another chalcogen element, $Y$, resulting in the formation of Janus TMDs ($M$ = Mo, W; and $\{X, Y\}$ = S, Se, Te; where $X \neq Y$) that stabilize in a hexagonal crystal structure as shown in Fig. 3(c) [88]. Typical Janus $MXY$ semiconductors have a stable structure at ambient conditions similar to the conventional 2D TMDs and possess intrinsic RSOC due to the built-in electric field present, perpendicular to the monolayer plane from the chalcogen atom with a larger atomic number (lower electronegativity) to the one with smaller atomic no (higher electronegativity) [89–91]. With the same chalcogenides, W$XY$ has a higher value of $\alpha_R$ than that of Mo$XY$ since W has a higher SOC than Mo. The value of $\alpha_R$ in these materials can be increased further with the application of an external electric field ($\mathbf{E}_{ext}$) parallel to the intrinsic field as it increases the charge accumulation in the chalcogen atom with a smaller atomic number [91]. Conversely, $\mathbf{E}_{ext}$ applied opposite to the intrinsic field weakens the Rashba effect. As presented in Fig. 4(a), among all other $MXY$ monolayers, WSeTe exhibits the most significant change in $\alpha_R$ with an increase of 0.031 eV Å near the $\Gamma$ point under an electric field of 0.5 V/Å when compared to its intrinsic value. On the contrary, MoSTe and WSTe are nearly insensitive to the applied field [91].

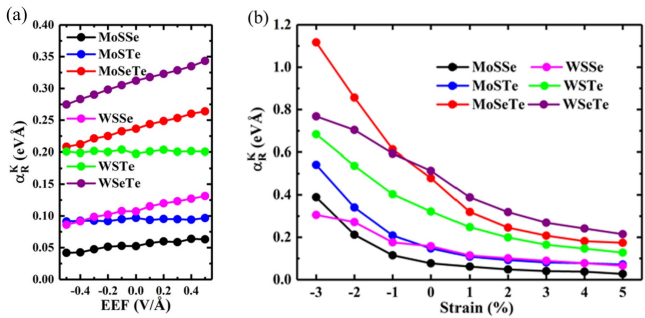

FIG. 4. (Color online) Modulation of the Rashba parameter $\alpha_R$ in 2D Janus TMDs under the influence of (a) an external electric field and (b) biaxial strain. Figure adopted from Ref. [91].

Janus TMDs exhibit a nonlinear change in $\alpha_R$ when subjected to an in-plane biaxial strain. Figure 4(b) depicts the change in $\alpha_R$ of the Janus MXY monolayers with applied biaxial strain. Among all others, MoSeTe shows the highest increase in $\alpha_R$ from 0.5 to 1.1 eV Å under a compressive strain of 3%. Hu et al. [91] also reported a significant increase in the anisotropic Rashba spin splitting of these materials with the application of a compressive strain.

On the other hand, compared to other Janus TMDs such as MoSSe and WSSe, strain-free monolayers of CrSSe, CrSTe, and CrSeTe show relatively higher values of the intrinsic Rashba parameter of 0.26, 0.31, and 1.23 eV Å, respectively. With an applied compressive strain of 2%, $\alpha_R$ of CrSSe, CrSTe, and CrSeTe monolayers increases up to 0.66, 0.50, and 2.11 eV Å, respectively [92]. Anisotropic Rashba spin splitting arising from RSOC is observed around the $M$-point in Pt$XY$ ($X = Y =$ S, Se, Te; $X \neq Y$) monolayers. The calculated values of $\alpha_R$ from the $M$- to the $\Gamma$-point are 1.654, 1.103, and 0.435 eV Å, while the values from the $M$- to $K$-point are 1.333, 1.244, and 0.746 eV Å for PtSSe, PtSTe, and PtSeTe, respectively [93].

RSOC can also be induced in centrosymmetric TMDs by applying an external electric field. $\mathbf{E}_{ext}$ breaks the inversion symmetry in these crystals and induces Rashba spin splitting. Yao et al. [94] reported a significant linear increase of $\alpha_R$ with increasing $\mathbf{E}_{ext}$ in six $MX_2$ monolayers. Figure 5 illustrates the variation in the Rashba parameter of 2D TMDs in response to an applied external electric field. Notably, the anions play a significant role in the electric-field dependence of RSOC in these materials. Conversely, the cations make a minimal contribution to the field dependence of RSOC as they are strongly shielded from the external electric field by the anions [94].

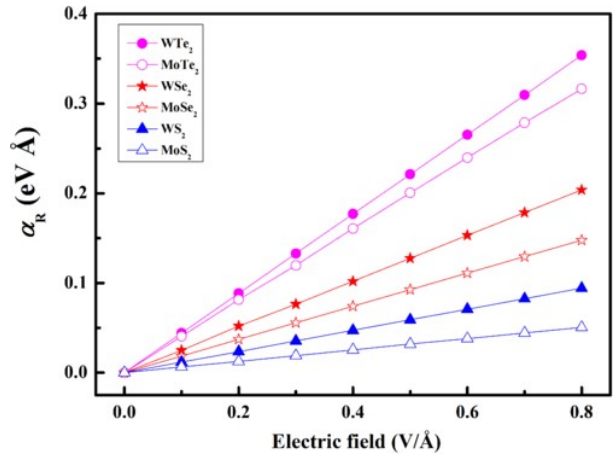

FIG. 5. (Color online) Variation in the Rashba parameter of 2D TMDs in response to an applied external electric field. Figure adopted from Ref. [94].

### C. Other 2D Janus monolayers

Apart from the Janus TMDs, various other Janus monolayers have also gained significant attention due to their large Rashba spin splitting and potential applications in spintronic devices. Bhat et al. [95] performed ab-initio calculations on Janus Sn$XY$ monolayers ($\{X, Y\} =$ S, Se, Te; with $X \neq Y$) to study how RSOC is affected by the replacement of the transition metal in Janus TMDs with a metal. The resulting Janus SnSSe, SnSeTe, and SnSTe monolayers exhibit anisotropic Rashba spin splitting having $\alpha_R$ of 0.109 eV Å, 0.273 eV Å, and 0.755 eV Å, respectively [95]. Notably, Bafekry et al. [96] reported a huge Rashba spin splitting in BiTeCl and BiTeBr monolayers with $\alpha_R$ as 7.48 eV Å and 9.15 eV Å, respectively. Conversely, BiTeI and SbTeI monolayers are reported to exhibit a relatively smaller $\alpha_R$ of 1.97 eV Å and 1.39 eV Å respectively [97]. The value of $\alpha_R$ for InTeF was determined to be 1.08 eV Å [98].

Another interesting class of 2D Rashba materials is the Janus transitional-metal trichalcogenide (TMTC) monolayers with the chemical formula of $MX_2Y$ ($M =$ Ti, Zr, Hf; $X \neq Y =$ S, Se). These materials show RSOC due to the cumulative effect of SOC and the lack of in-plane mirror symmetry. In the case of $M$S$_3$-based TMTC monolayers, the Rashba spin splitting occurring near the $\Gamma$ point of VBM is mainly contributed by the $d_{xy}$ orbitals of the transition-metal atoms. The Rashba parameter in these systems varies as $TiS_2Se > ZrS_2Se > HfS_2Se$ due to decrease in the built-in electric field. On the contrary, $MSe_3$-based TMTC monolayers do not exhibit any RSOC, owing to the lack of any contribution from the $d_{xy}$ orbitals of the transition metal near the Fermi level. The calculated value of $\Delta E_R$, $\Delta k_R$ and $\alpha_R$ for the Janus $TiS_2Se$ monolayer is 40 meV, 0.074 Å$^{-1}$, and 1.081 eV Å, respectively [99].

Indirect band gap semiconductors of 2D Janus tellurite ($Te_2Se$) monolayers show RSOC near the $\Gamma$ point of CBM, the significant contribution of which arises from the $p_z$ orbitals of Te atoms [100]. Guo et al. [101] recently predicted a new class of Janus materials $MA_2Z_4$ ($M$ = transition metal such as W, V, Nb, Ta, Ti, Zr, Hf, or Cr; $A$ = Si or Ge, and $Z$ = N, P, or As) with intrinsic RSOC. The predicted monolayers of $MSiGeN_4$ ($M$ = Mo and W) are dynamically and thermally stable and exhibit RSOC with a Rashba energy of 0.8 meV and 4.2 meV for $MoSiGeN_4$ and $WSiGeN_4$, respectively [101, 102].

An exciting class of 2D Janus Rashba materials with a substantially large value of $\alpha_R$ has been recently proposed by Karmakar et al. [66]. These materials belong to the Janus $Mo_2COX$ structures, which are derived from the parent compound $Mo_2CO_2$, a popular class of 2D MXenes in the family of 2D transition metal carbides/nitrides/carbonitrides with the generic formula of $M_{n+1}X_nT_2$ ($M$ = 3$d$ or 4$d$ transition metals, $X$ = C or N, $T$ = surface termination unit, $n$ = 1–3). Here, RSOC is induced by breaking the inversion symmetry present in the parent compound $Mo_2CO_2$, with the replacement of one of the terminating O layers either with a halogen (F, Cl, Br, and I) or a chalcogen (S, Se, and Te) layer [see Fig. 3(d)].

### D. Rashba effect at interfaces and junctions

Constructing vdW heterostructures is a promising approach for manipulating the properties of the constituent monolayers. The structural asymmetry within these heterostructures creates a net out-of-plane electric field which impacts the system's RSOC. Consequently, electronic properties undergo significant changes compared to the individual monolayers. For instance, with the formation of a bilayer, Liu et al. [85] observed an almost two-fold increase in the Rashba spin splitting at the CBM of $M$Te ($M$ = Ge, Sn, and Pb) compared to its monolayer. This increase is attributed to the doubling of the dipole moment. The calculated values of $\alpha_R$ are 1.10, 1.02, and 1.05 eV Å for GeTe, SnTe, and PbTe bilayers, respectively [85].

The 2D III-VI chalcogenide $NX$ ($N$ = Ga, In; $X$ = S, Se, Te) monolayers do not exhibit any intrinsic RSOC owing to the out-of-plane mirror symmetry. Nevertheless, this limitation is overcome by constructing vdW heterostructures of InSe with GaTe and InTe. These heterostructures, namely InSe/GaTe and InSe/InTe, exhibit significant RSOC at the CBM with $\alpha_R$ of 0.50, and 0.44 eV Å, respectively, where RSOC mainly originates from the InSe layer [104]. First principles studies on the Ga$X$/$MX_2$ ($M$ = Mo, W; $X$ = S, Se, Te) heterostructures reveal an increase in Rashba spin splitting with increased SOC in the $p$-orbitals of chalcogen atoms. Strikingly, the replacement of Mo with W decreases the strength of RSOC due to the increase in $d$-orbital contribution near the valance band edge at the expense of the $p$-orbital of the chalcogen atoms confirming the major contribution of the $p$-orbitals to the Rashba spin splitting. Similar observations hold true for the $MoS_2/Bi(111)$ system [105].

Furthermore, InTeF/AlN and InTeF/BN heterostructures exhibit $\alpha_R$ of 1.13 and 1.08 eV Å, respectively [98]. A significant Rashba spin splitting with an energy of 110 meV was observed near the $\Gamma$ point of $PtSe_2/MoSe_2$ heterostructures, which is mainly attributed to the strong interfacial SOC arising from the hybridization between the two constituent monolayers [106]. Moreover, by applying strain and electric fields, the RSOC in these materials can be tuned effectively. Particularly, with the application of 6 % in-plane strain, the Rashba coefficient increases up to 1.33 and 1.26 eV Å for InSe/GaTe and InSe/InTe, respectively [104]. The first-principles calculations on the RSOC exhibited by InTe/$PtSe_2$ heterostructures reveal Rashba-type spin splitting near the $\Gamma$ point of VBM, which can be manipulated by strain engineering. While tensile strain increases the $\alpha_R$, compressive strain weakens the RSOC. On the other hand, strong interlayer coupling promotes RSOC resulting in an enhanced $\alpha_R$ with decreased interlayer distances [107].

Peng et al. [108] report a significant Rashba spin splitting near the $\Gamma$ point of $MoS_2/Bi_2Te_3$ that mainly arises from the $Bi_2Te_3$ layer, while the $MoS_2$ layer plays an inductive role. The $\alpha_R$ in these systems are comparable to those observed in $\{Bi_2Se_3\}_2/InP(111)$ heterostructures as observed from the angle-resolved photoemission spectroscopy (ARPES) measurements. The spin textures measured experimentally for different thicknesses of $\{Bi_2Se_3\}_2/InP(111)$ heterostructures reveal Rashba-like splitting of the massive Dirac cones in the surface states as a result of the substrate-induced inversion asymmetry [100].

In a recent study, Sattar et al. [109] observed that $\alpha_R$ varies significantly with changing thickness of the constituting layers in $Bi_2Se_3/PtSe_2$ heterostructures. Specifically, the heterostructure consisting of 2-quintuple layers (QL) $Bi_2Se_3$/ 2L $PtSe_2$ exhibits $\alpha_R$ of 16.84 eV Å and 4.4 eV Å at VBM and CBM respectively, which are among the highest reported values for known 2D Rashba materials [109]. Intrinsic RSOC is also present in Janus $SnSSe/WSSe$ semiconductor heterostructures, which exhibit a Rashba spin splitting near the Fermi level with $\alpha_R$ of 0.7 eV Å. With the application of 12.8 % of compressive strain this $\alpha_R$ increases up to 1.07 eV Å. On the contrary, metallic heterostructures of Janus layers SnSTe/WSTe and SnSeTe/WSeTe do not exhibit any Rashba-type splitting [95].

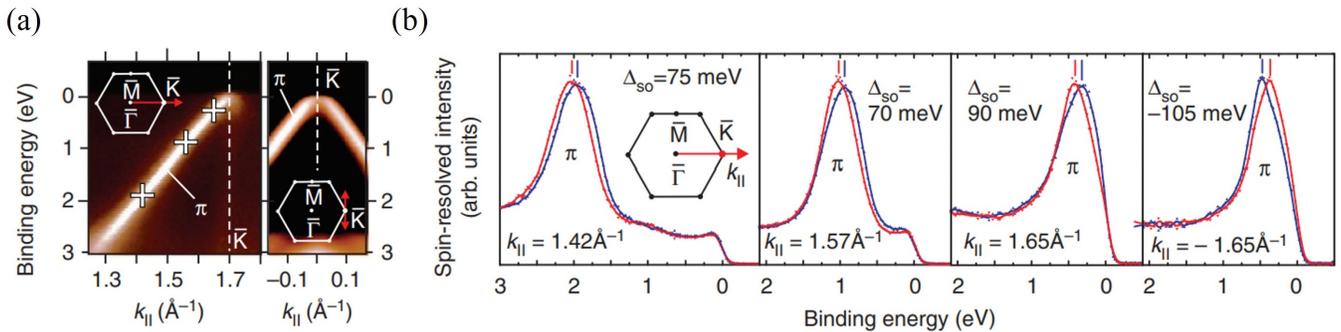

FIG. 6. (Color online) (a) ARPES spectra around the K point in the Brillouin zone of graphene in graphene/Au/Ni(111). Cross symbols indicate the positions in the Dirac cone of the $p$-states in graphene probed by spin- and angle-resolved photoemission spectra as shown in (b). Blue (red) lines in (b) represent spin-up (down) spectra. $\mathbf{k}_{||}$ is the in-plane component of the wave vector. Figure adopted from Ref. [103].

Further, Kunihashi et al. [110] studied the effect of incorporating heavier Bi elements into GaAs heterostructures by employing time- and spatially- resolved Kerr rotation measurements on the 70 nm thick epitaxially grown layer of $GaAs_{0.961}Bi_{0.039}$. Their results revealed that RSOC prevailed over the Dresselhaus effect, with the measured value of $\alpha_R$ as 2.5 meVÅ. Interestingly, a decrease in $\alpha_R$ was observed with an increase in the pump laser intensity due to the light-induced screening effect on the built-in potential gradient [110]. GaSe/MoSe$_2$ heterostructures also exhibit a Rashba splitting of 0.49 eV Å. Notably, these heterostructures exhibit the coexistence of Rashba splitting and band splitting at the $K$ and $K'$ valleys, highlighting their potential for applications in spintronics and valleytronics [111].

### E. Proximity-induced RSOC

Proximity-induced SOC primarily occurs within a few atomic layers of the low-SOC material near its interface when a heterostructure is formed with a material exhibiting stronger SOC. This phenomenon may result in the charge transfer between the material and the substrate or the formation of a thin interfacial layer with slightly altered electronic states [113–116]. Graphene shows a substantial increase in RSOC with the proximity effect of substrates. In its free-standing form, graphene possesses the intrinsic spin-orbit splitting of around 50 µeV. However, the Rashba spin splitting of the Dirac cones can be enhanced up to 13-100 meV when synthesized epitaxially with high SOC elements including Ag, Au, and Pb intercalated, or in direct contact with different substrates such as Ni, Ir, and Co [103, 117–122]. The anomalous increase in the spin splitting can be attributed to a strong π-d hybridization between graphene and the substrate. Figures 6(b) shows the variation in RSOC of graphene in graphene/Au/Ni(111) heterostructure on account of the proximity-induced effect of the substrate.

The use of thin layers of 3D topological insulators like Bi$_2$Se$_3$ or Bi$_2$Te$_3$ as substrate can further increase RSOC in graphene [123, 124]. On the other hand, TMDs, as mentioned above, do not exhibit any intrinsic RSOC. However, MoTe$_2$ when placed on top of EuO substrate shows a relatively large Rashba spin splitting owing to the proximity-induced interactions [125]. Furthermore, the Janus WSSe semiconductor exhibits an increase in $\alpha_R$ from 0.17 to 0.95 eVÅ under the proximity effect of MnO (111) [126].

### F. Nontrivial topological phase induced by RSOC

Recent theoretical and experimental studies on topological insulators (TI) have developed a profound understanding of how electronic band structure in certain materials can be altered significantly by SOC effects. SOC often leads to band inversion near the Fermi level, driving the topological phase transition in materials [69–72]. In the case of 2D (3D) TIs, this type of transformation leads to the formation of Dirac-cone states, characterized by a distinctive spin-momentum interdependence on the metallic edge (surface) states. This exciting feature forms the basis of the quantum spin Hall effect, with profound implications for condensed matter physics and potential applications in novel spintronic devices. Due to the conservation of TR-symmetry, the opposite edge states of the 2D nontrivial TIs possess opposite spin chirality and are charge neutral [127–130].

BiTeI well exemplifies this phenomenon. Under normal conditions, BiTeI exhibits a substantial Rashba spin splitting of 3.85 eV Å. Strikingly, BiTeI can transform a topologically trivial state to a nontrivial TI by applying hydrostatic pressure, as shown in Figure 7. Under ambient conditions the system possesses a conventional band gap of 286 meV and a Rashba energy of 110 meV. However, with hydrostatic compression, beyond a crit-ical point an inverted band gap emerges between the top valence band and the bottom conduction band leading to a topologically nontrivial phase [112]. Similarly, Jozwiak et al. [131], based on their DFT calculations on a 7-quintuple layer thick slab of $Bi_2Se_3$, demonstrated that the emergence of surface band inversion in the surface electronic configuration of the topological insulator $Bi_2Se_3$ is mainly caused by RSOC [131–133].

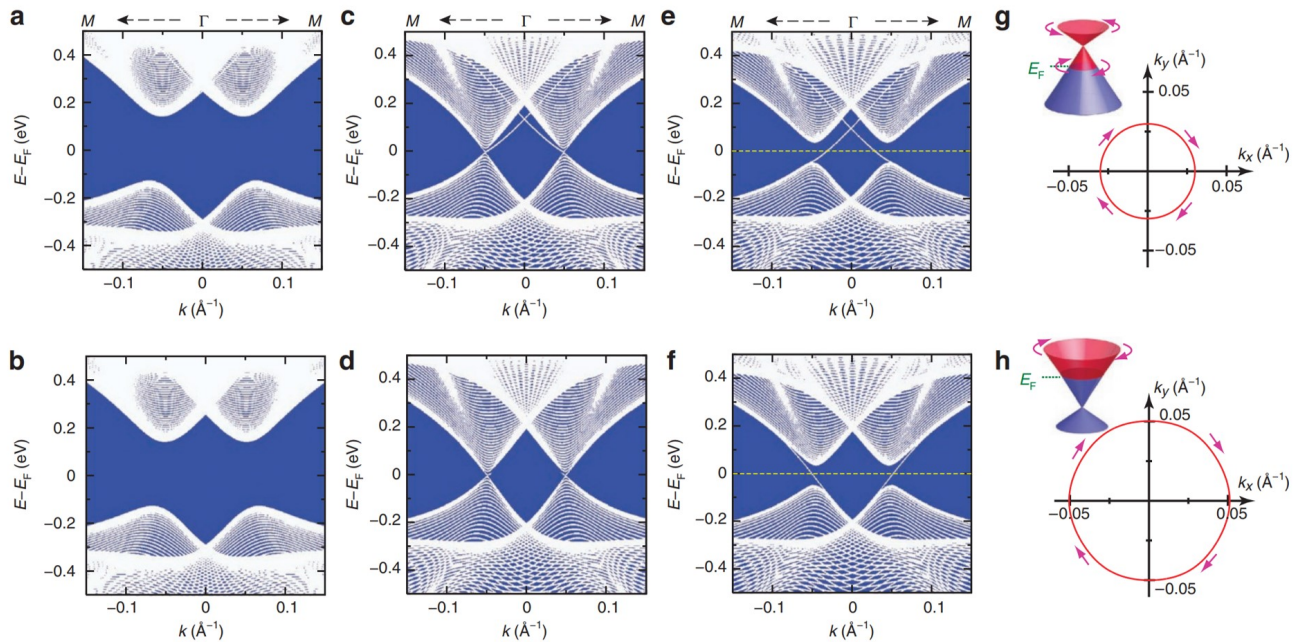

FIG. 7. (Color online) Computed electronic band dispersions in proximity to the Fermi level (EF) for both I-terminated (top panels) and Te-terminated (bottom panels) surfaces of BiTeI, subjected to hydrostatic compression by (a, b) V/V0=1, (c, d) V/V0=0.89, and (e, f) V/V0=0.86. The corresponding Fermi surfaces for cases (e) and (f) are depicted in (g) and (h) respectively. Figure adopted from Ref. [112].

In 2D materials, one of the intriguing materials featuring the coexistence of both RSOC and topologically nontrivial edge states is the Janus RbKNaBi monolayer [134]. It is a quantum spin hall insulator with a relatively large band gap and is dynamically and thermally stable. RbKNaBi possesses intrinsic RSOC owing to a built-in electric field because of the difference in electronegativities between the top and bottom atomic layers. Figures 8(a) and 8(b) show the electronic band structure of RbKNaBi with the orbital projection of $s$, $p_x + ip_y$, and $p_z$ orbitals of Bi, calculated using GGA and GGA + SOC approximations, respectively. A SOC-induced band inversion can be noticed at the Fermi level near the Γ point. Interestingly, RbKNaBi shows a topologically nontrivial behavior, as shown in Figs. 8(c,d). A single pair of helical edge states are present within the band gap of nearly 229 meV. The band gap is sufficiently large to protect the helical edge states against thermally activated carriers, enabling the realization of the room temperature quantum spin Hall effect.

An intriguing area of research has emerged recently, focusing on proposing artificial engineering of TIs through strategic layering of topologically trivial Rashba monolayers, utilizing first-principles calculations. In 2013, Das et al. [135] introduced a novel approach of designing a 3D TI by stacking bilayers composed of two-dimensional Fermi gases, each exhibiting opposite RSOC on adjacent layers. They observed that while a single bilayer consistently demonstrated topologically trivial behavior, topologically nontrivial insulating states emerged only in the bulk after crossing a critical number of bilayers. On the other hand, Nechaev et al. [136], through theoretical investigations, found that a centrosymmetric sextuple layer formed by combining two BiTeI trilayers with opposite RSOC exhibits an inverted bandgap of sufficient magnitude for practical applications. However, they observed that the sextuple layer transitioned to a topologically trivial state with just a 5% increase in vdW spacing. This strategic approach presents new avenues for designing intriguing topological materials by leveraging the inherent RSOC in 2D materials.

### G. Optical manipulation of Rashba effect

RSOC plays a crucial role in spin field-effect transistors, enabling information processing and storage without reliance on external magnetic fields. However, the spin relaxation mechanism in the 2D semiconducting channels limits the precision and accuracy of these devices. Achieving a persistent spin helix (PSH) condition in the 2DEG is an effective solution to tackle this chal-lenge. Hence, efficient engineering of RSOC and Dresselhaus effect in 2D materials is of subsequent importance [137]. While the Rashba parameter can be manipulated by an external electric field and/or strain, and the Dresselhaus effect can be tuned by controlling the width of the quantum wells [138], these adjustments often require complex fabrication processes. However, replacing the gate in state-of-the-art s-FETs with an optical field presents a promising avenue for developing faster and more energy-efficient devices. This approach offers a flexible and efficient means to control the RSOC [139]. Notably, optical tuning is a non-destructive as well as reversible technique that can accurately alter the electron density and effectively screen the intrinsic electric field in the system without relying on the excitation beam [140].

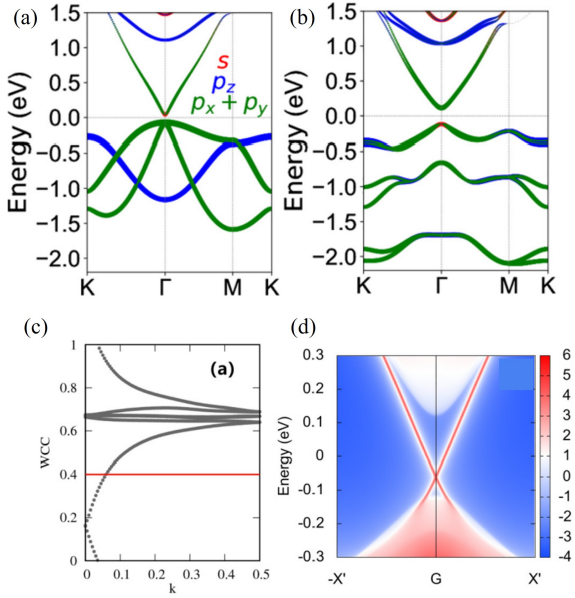

FIG. 8. Electronic band structure of RbKNaBi with the orbital projection of s, $p_x + p_y$, and $p_z$ orbitals of Bi calculated (a) without and (b) with SOC. (c) Evolution of WCC computed using GGA + SOC, indicating a nontrivial $Z_2$ topological invariant ($Z_2 = 1$). (c) Projected edge spectrum revealing a nontrivial metallic edge state within the bandgap. Figure adopted from Ref. [134].

Ma *et al.* [141] investigated the optical tuning of RSOC and Dresselhaus effect in the 2DEG of $GaAs/Al_{0.3}Ga_{0.7}As$ by measuring the spin-galvanic effect (SGE). They introduced an additional control light above the barrier's bandgap to tune the SGE excited by a circularly polarized light below the bandgap of GaAs. Their observations reveal an efficient optical tunability of RSOC compared to Dresselhaus SOC in $GaAs/Al_{0.3}Ga_{0.7}As$. Specifically, they demonstrate that the ratio of Rashba- and Dresselhaus-related SGE currents varies systematically with the increasing power of the control light, as shown in Fig. 9(a). Above a critical point inverse PSH emerges of resulting an extended spin lifetime. This emphasizes the potential of optical tuning as an effective technique for modulating SOC, offering implications for the design of spintronic devices with prolonged spin coherence time [141].

On the other hand, Michiardi *et al.* [142], as part of their study on topological insulator $Bi_2Se_3$, demonstrated efficient tuning of RSOC using optical pulses with a picosecond timescale. According to their proposed mechanism, optical excitation above the energy gap induces charge redistribution perpendicular to the surface in the presence of a band-bending surface potential. This generates an ultrafast photovoltage that modulates the $\alpha_R$ within a sub-picosecond timeframe. The measured change in Rashba momentum $\Delta k_R$ within the first quantum well state (QWS1), approximately $3.5 \times 10^{-3}\,\text{Å}^{-1}$, corresponds to an alteration in the spin precession angle of $\pi$ over a distance of less than 100 nm [142]. This implies that the effect becomes noticeable in devices of similar length under ballistic transport conditions. The use of optical pulses to manipulate Rashba splitting in 2DEGs with a picosecond-level timescale represents a significant advancement in optically controlled spintronic devices.

Another study on $Ge/Si_{0.15}Ge_{0.85}$ multiple quantum wells further supports the viability of contactless optical excitation as an effective method for tuning SOC, thus paving the way for electro-optic modulation of spin-based quantum devices consisting of group IV heterostructures [143].

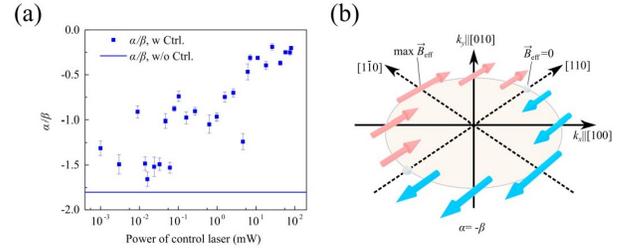

FIG. 9. (a) Optical manipulation of the Rashba-to-Dresselhaus coefficient ratio ($\alpha/\beta$) in $GaAs/Al_{0.3}Ga_{0.7}As$ with increasing power of the control light, displayed in a semilogarithmic plot. The solid line illustrates the baseline ratio in the absence of control light. (b) Orientation of the effective magnetic field vector, $\mathbf{B}_{eff}$, within the momentum space when the inverse persistent spin helix ($\alpha = -\beta$) condition is satisfied. While the arrows indicate the direction of $B_{eff}$, their lengths symbolize the corresponding field strength. $B_{eff}$ is the resultant of the Dresselhaus and Rashba effective magnetic fields. A unidirectional alignment of $B_{eff}$ is ensured by the inverse persistent spin helix condition. Figure adopted from Ref. [141].

### H. Electrically-controlled 2D Rashba systems

More recently, the realization of 2D ferroelectricity has been demonstrated in various compounds [144–151]. Some of which, profit from the spontaneous polarization coupling with the strong SOC leading to a reversible Rashba 2D compound. In striking difference to bulk ferroelectrics that aim to be growth in ultra-thin films, the 2D ferroelectric materials do not suffer from the polarization cancellation induced by the depolarizing fields. The latter presented as a result of the ultra thin-films dimensionality. Additionally, due to the weak vdW interaction between the layers, the 2D ferroelectric Rashba compounds are ideal for the experimental processing showing a large CMOS growth compatibility. Finally, and as an additional advantage of the 2D ferroelectric Rashba compounds is that, as demonstrated in the WTe$_2$ compounds [152, 153], the covalent bonding in the layer favors the charge screening resulting in a metallic ferroelectricty behavior ideal for controlling the spin-texture by electric field in 2D devices. For more details on 2D-ferroelectric Rashba semiconductors (2D-FERSC), we refer the reader to Refs. [144, 145, 150].

TABLE I: List of Rashba parameters ($E_R$, $k_R$, and $\alpha_R$) of some 2D Rashba materials. Additionally, here are displayed several tentative functionalities depending on the family and particular compound. Those could cover, but are not limited to, spin field-effect transistors (s-FET), 2D-ferroelectric Rashba semiconductors (2D-FERSC), and optical-Rashba devices (ORD).

| Materials | $E_R$ (meV) | $k_R$ (Å$^{-1}$) | $\alpha_R$ (eVÅ) | Ref. | Functionality |
|---|---|---|---|---|---|
| **AB binary monolayer** | | | | | |
| h-TaN | 74 | - | 4.23 | [82] | |
| h-NbN | 52 | - | 2.9 | [82] | |
| AlBi | 22 | 0.016 | 2.8 | [154, 155] | s-FET, 2D-FERSC |
| PbSi | 9 | 0.007 | 2.7 | [154, 155] | s-FET, 2D-FERSC |
| BiSb | 13 | 0.011 | 2.3 | [64] | |
| BiAs | - | - | 1.95 | [154] | s-FET |
| TlP | - | - | 1.79 | [154, 155] | s-FET, 2D-FERSC |
| BiP | - | - | 1.6 | [154] | |
| PbBi | 20 | 0.025 | 1.6 | [80] | |
| PBi | - | - | 1.56 | [80] | |
| GaSb | - | - | 1.45 | [155] | s-FET, 2D-FERSC |
| BiN | - | - | 1.31 | [154] | |
| WSe | - | - | 1.26 | [86] | |
| WS | - | - | 1.2 | [86] | |
| BiB | - | - | 1.16 | [154, 155] | s-FET |
| ZnTe | - | - | 1.06 | [156] | |
| PbTe bilayer | - | - | 1.05 | [85] | |
| SnTe bilayer | - | - | 1.02 | [85] | |
| WC | - | - | 1.02 | [86] | |
| GeTe bilayer | - | - | 1.0 | [85] | s-FET, 2D-FERSC |
| GaTe | 15 | 0.029 | 1.0 | [87] | |
| MgTe | - | - | 0.63 | [83] | |
| GeTe monolayer | 2 | - | 0.6 | [85] | |
| PbTe monolayer | 2 | - | 0.6 | [85] | |
| SnTe Monolayer | - | - | 0.6 | [85] | |
| PSb | - | 0.006 | 0.4 | [80] | |
| MoC | - | - | 0.14 | [86] | |
| PAs | 0.1 | 0.002 | 0.1 | [80] | |
| **2D Janus Monolayers** | | | | | |
| BiTeBr | - | - | 9.15 | [96, 157] | s-FET |
| BiTeCl | - | - | 7.48 | [96] | |
| Mo$_2$COI (AA) | 0.1129 | 0.0571 | 3.9491 | [66] | |
| Mo$_2$COTe (BB) | 0.0768 | 0.0640 | 2.3967 | [66] | |
| Mo$_2$COSe (BB) | 0.1789 | 0.1470 | 2.3247 | [66] | |
| BiTeI | 40 | 0.043 | 1.97 | [97, 157] | s-FET |
| Mo$_2$COS (BB) | 0.1789 | 0.1879 | 1.9045 | [66] | |
| Janus Sb$_2$Se$_2$Te armchair (zigzag) | - | - | 1.53 (1.52) | [158] | |
| Mo$_2$COCl (BB) | 0.1022 | 0.1376 | 1.4854 | [66] | |
| SbTeI | 17 | 0.024 | 1.39 | [97] | |
| CrSeTe | - | - | 1.23 | [92, 155] | |
| TiS$_2$Se | 40 | 0.074 | 1.081 | [99] | |
| InTeF monolayer | - | - | 1.08 | [98] | |
| Sb$_2$SeTe$_2$ monolayers: armchair (zigzag) | - | - | 1.00 (1.12) | [158] | |
| WSeTe | 52 | 0.17 | 0.92 | [155, 159, 160] | s-FET |
| Mo$_2$COBr (AA) | 0.0072 | 0.0176 | 0.8185 | [66] | |
| SnSTe | 6.95 | 0.0184 | 0.755 | [95] | |
| WSSe | 3.6 | 0.01 | 0.72 | [155, 159] | |
| ZrS$_2$Se | 19 | 0.053 | 0.717 | [99] | |
| HfS$_2$Se | 15 | 0.053 | 0.566 | [99] | |
| MoSSe | 1.4 | 0.005 | 0.53 | [155, 159] | s-FET |
| CrSTe | - | - | 0.31 | [92] | |
| RbKNaBi | 1.3 | - | 0.274 | [134] | |
| SnSeTe | 2.46 | 0.018 | 0.273 | [95] | |
| CrSSe | - | - | 0.26 | [92, 155] | |
| WSTe | 7.78 | 0.0631 | 0.247 | [95, 155] | s-FET |
| WSiGeN$_4$ | 4.2 | 0.076 | 0.111 | [101] | |
| SnSSe ($\Gamma$-$K$) | 1.03 | 0.0189 | 0.109 | [95] | |
| MoSiGeN$_4$ | 0.8 | 0.048 | 0.033 | [101] | |
| MoSeTe | - | - | 0.012 | [155, 159] | s-FET |
| **2D Van der Waals Hetrostructures** | | | | | |
| 1QL(2QL) Bi$_2$Se$_3$/1L PtSe$_2$ | 4.8 (4.0) | 0.002 (0.002) | 4.8 (4.0) | [109] | |
| BiSb/AlN | - | - | 1.5 | [64] | |
| PtSe2/MoSe$_2$ | - | - | 1.3 | [106] | s-FET, ORD |
| AlN/InTeF | 11 | - | 1.13 | [98] | |
| BN/InTeF | 10 | - | 1.08 | [98] | |
| J-SnSSe/WSSe | 42.91 | 0.126 | 0.681 | [95] | |
| Bi(111) surface | 14 | 0.05 | 0.55 | [161] | |
| InSe/GaTe | - | - | 0.5 | [104] | |
| GaSe/MoSe$_2$ | 31 | 0.13 | 0.49 | [162] | s-FET |
| InSe/InTe | - | - | 0.44 | [104] | |
| Au (111) surface | 2.1 | 0.012 | 0.33 | [163] | |
| J-SnSTe/WSTe | 2.47 | 0.0366 | 0.135 | [95] | |
| InGaAs/InAlAs surface | <1.0 | 0.028 | 0.07 | [164, 165] | s-FET |
| LaAlO$_3$/SrTiO$_3$ interface | ~1.0 | - | 0.02 | [166, 167] | ORD |
| GaS/MoS$_2$ | 2 | 0.05 | - | [162] | s-FET |
| GaS/WS$_2$ | 1 | 0.03 | - | [162] | s-FET, ORD |
| GaSe/WSe$_2$ | 22 | 0.11 | - | [162] | s-FET |
| GaTe/MoTe$_2$ | 48 | 0.12 | - | [162] | s-FET |
| GaTe/WTe$_2$ | 47 | 0.11 | - | [162] | s-FET |
| KTaO$_3$/K(Zn,Ni)F$_3$ interface | 64 | 0.044 | - | [167, 168] | ORD |

## IV. APPLICATIONS IN SPINTRONIC INDUSTRY: POSSIBLE DEVICE REALIZATIONS

Spintronic presents a promising next-generation platform, surpassing traditional electronics by leveraging electrons' spin degree of freedom [7–9, 169, 170]. Furthermore, manipulating spin as a logical unit in spintronic opens up novel avenues for neuromorphic [171, 172] and probabilistic computing [173]. Spintronic improves scaling, processing speed, and energy efficiency compared to electronics and establishes a direct interface with existing technologies. For instance, spin valve and magnetic tunnel junction (MTJ) devices found swift applications in disk read heads, proximity sensors for automobiles, automated industrial tools, and biomedical devices following the discovery of giant magnetoresistance [174, 175]. However, reliance on external magnetic field limits the energy efficiency of MTJ- and GMR-based devices.

The discovery of spin-transfer torques (STT) [176, 177] enables all-electric control over spin states and resistance in GMR devices, enhancing scalability. This has led to the development of scalable nonvolatile magnetic random access memory (RAM) using STT, replacing static RAM and showing potential applications in dynamic RAM technology [178–180]. While GMR devices are currently integrated into conventional electronic platforms, an all-spintronic platform requires further innovation in materials design and the fabrication of high-density and low-power components. Notably, STT devices face challenges such as dependence on high-performance magnets, spin filtration, low spin carrier lifetimes and diffusion lengths, Joule heating, and voltage breakdown [7–9, 169].

Recent phenomena include the spin Hall effect (SHE) [181–185] and Rashba-Edelstein effect (REE) [182], along with their optical [186] and thermal [187] equivalents, introduce new possibilities for efficient spin manipulation. Focusing on SHE and REE, an applied voltage generates a spin-polarized current and interfacial spin accumulation, similar to STT but without charge flowing through the magnetic layers. This process reduces the impact of Joule heating and minimizes the risk of voltage breakdown compared to STT. Additionally, these spin-orbit torques (SOT) can excite various magnetic materials [188], enabling the switching of single-layer magnets and efficient manipulation and excitation of domain walls, skyrmions, and spin waves. However, progress with SOT is hindered by low charge-spin conversion efficiency [7] and reliance on heavy elements.

A diverse range of spin transistors, leveraging various operational principles, has been proposed in the literature [170, 189]. One notable example demonstrating practical applications of RSOC is the Datta-Das spin transistor, also referred to as the spin FET [47]. In this transistor, an external gate voltage controls the flow of spin-polarized electrons. The device typically consists of a Rashba semiconductor (channel) with two ferromagnetic contacts, which act as a source and drain (see Fig. 1). The gate voltage is used to manipulate the spin of electrons, allowing for the modulation of the spin current between the ferromagnetic contacts.

Furthermore, the concept of a bipolar spin switch, introduced by Johnson [190] in 1993, outlines a spin injection technique employing a thin ferromagnetic film to polarize the spin axes of electrons transporting an electric current in a ferromagnetic-nonmagnetic-ferromagnetic metal trilayer structure. This configuration yields a three-terminal, current-biased device with a bipolar voltage (or current) output dependent on the magnetization orientations of the two ferromagnets.

As observed in layered bulk compounds [191–195], the intrinsic coupling between the broken inversion symmetry and SOC opens the possibility of an electrically-controlled Rashba device. In this potential device design, realization relies on the feasibility of polarization switching in the 2D ferroelectric layer, unlocking spin-texture reversal and spin control. Potential candidates include the {Mo,W}Te$_2$ [152, 153, 196–198], AgBiP$_2$Se$_6$ [144], and AB monolayers [144] in which, the out-of-plane polarization can be switched. Some efforts towards such direction have shown the successful design of the electric field control of valleytronics [155, 199, 200]. Moreover, an additional advantage of the 2D vdW ferroelectrics is the feasible functionalization and growth on substrates, offering a doable engineering of the potential device.

Other proposed devices profit from the Aharonov–Casher effect [201] in which charge-neutral magnetic moments experience quantum oscillations in the presence of an external electric field [202]. Such proposal considers Rashba active materials ring shaped, by lithography for examples, in which, the spin momenta, associated to the electons's flow, present a difference at the end of the loop.

## V. SUMMARY AND OUTLOOK

Highlighted in this Perspective is the crucial significance of the Rashba effect in broadening the research horizons within the domain of 2D materials, surpassing the confines of graphene. This investigation encompasses transition-metal dichalcogenides, silicene, germanene, and stanene. The Rashba effect enhances the comprehension of these materials and establishes the groundwork for actualizing diverse unprecedented physical occurrences and technological advancements. Key areas influenced by the Rashba effect include:

**Spin-Orbitronics:** This emerging field represents a synergistic blend of spintronics and orbitronics, where both spin and orbital degrees of freedom are manipulated [203, 204]. The Rashba effect plays a crucial role in this integration, enabling the simultaneous control of spin and orbital characteristics, which could revolutionize the design and functionality of electronic devices [205].

**Nonlinear Spintronics:** In strong electron-electron interactions, the Rashba effect contributes to the emergence of nonlinear spintronic phenomena [206, 207]. This interaction can lead to higher-order harmonics in spin currents, opening up new possibilities for advanced information processing and storage technologies that leverage these complex spin dynamics. Furthermore, strong electron-electron interactions in Rashba materials can yield non-conventional correlated states, unusual collective modes, bound electron pairs with non-trivial orbital and spin structures [73–78].

**Spin-Photovoltaics and Optospintronics:** This area explores the intersection between photonics and spintronics, where the Rashba effect facilitates the coupling between light and spin-polarized currents [208, 209]. This coupling could lead to novel spin-photovoltaic devices, which harness light to generate spin currents, and optospintronics, which combines optical and spintronic functionalities for innovative applications.

**Thermal Spintronics:** The Rashba effect is instrumental in controlling spin currents induced by thermal gradients, an essential aspect of thermal spintronics [210, 211]. This control is crucial for developing spin-based thermoelectric devices, which can convert waste heat into sound energy, offering a novel approach to energy efficiency and sustainability.

**Ultrafast Spin Dynamics:** The Rashba effect provides an ideal platform for the ultrafast manipulation of spin states [212]. This capability is essential for creating ultrafast memory devices, where rapid and precise control over spin dynamics is paramount and can use materials with a significant Rashba effect [213]. The ability to manipulate spin states on extremely short timescales could lead to a new generation of high-speed, high-efficiency memory and processing devices, significantly outperforming current technologies in speed and energy consumption.

Each of the aforementioned domains signifies a notable progression in materials science and technology, steered by the distinctive characteristics of the Rashba effect in 2D materials. The exploration of Rashba materials, particularly in the context of 2D vdW materials, has made significant strides. However, despite the extensive list of identified Rashba materials, there remains a gap in developing a comprehensive descriptor that can effectively pinpoint optimal systems with isolated spin states and large tunable splitting [62]. The existence of heavy atoms in these materials stands as a pivotal element owing to their robust SOC, a fundamental aspect in the emergence of the Rashba effect. Furthermore, 2D layers with pronounced crystal-potential gradients are imperative. Nevertheless, the obstacle resides in pinpointing explicit, measurable factors that can be methodically utilized to foresee and enhance Rashba attributes in the materials' design phase. This absence of a conclusive array of parameters or a descriptor constrains the capacity to effectively engineer materials with targeted Rashba characteristics.

From a theoretical perspective, current methodologies primarily involve analyzing the spin texture to confirm the existence of Rashba effects. While effective, this approach often requires a comprehensive analysis of the electronic band structure, including SOC effects, which can be complex and resource-intensive. A more streamlined method to calculate the Rashba parameter would be highly beneficial. Ideally, such a methodology would allow for estimating $\alpha_R$ without necessitating a complete band structure analysis, thereby simplifying the identification and characterization of Rashba materials in high-throughput calculations.

To address all these challenges, future research could focus on:

- *Developing Predictive Models*: Machine learning and data-driven approaches could be employed to develop predictive models that identify potential Rashba materials based on their atomic and electronic properties [62]. Even if regression models are trained, obtaining a reasonable estimation of the Rashba parameter would be possible.

- *High-Throughput Screening*: Leveraging computational tools for high-throughput screening of materials could expedite the discovery of new Rashba systems using established criteria and theoretical models [214, 215].

- *Advanced Computational Methods*: Improving computational methods to more efficiently calculate $\alpha_R$ (beyond the linear-**k** Rashba model) and other relevant parameters, possibly through developing new algorithms or adapting existing ones to target Rashba-related properties specifically.

- *Experimental Validation*: Complement theoretical advancements with experimental techniques to validate predictions and refine models, ensuring the theoretical descriptors are grounded in practical, observable phenomena.

In essence, although notable advancements have been achieved in pinpointing 2D Rashba materials, there exists a distinct requirement for more sophisticated descriptors and computational approaches to scrutinize both traditional and unconventional (hidden) Rashba systems. These progressions would substantially augment the capacity to devise and leverage materials with precise Rashba attributes, thereby facilitating a more effective and focused progression in spintronics development.

### ACKNOWLEDGEMENTS

AB and SS acknowledge support from the University Research Awards at the University of Rochester. SS is supported by the U.S. Department of Energy, Office of Science, Office of Fusion Energy Sciences, Quantum Information Science program under Award No. DE-SC-0020340. AHR thanks the Pittsburgh Supercomputer Center (Bridges2) and San Diego Supercomputer Center (Expanse) through allocation DMR140031 from the Advanced Cyberinfrastructure Coordination Ecosystem: Services & Support (ACCESS) program, which is supported by National Science Foundation grants #2138259, #2138286, #2138307, #2137603, and #2138296. AHR also recognizes the support of West Virginia Research under the call Research Challenge Grand Program 2022 and NASA EPSCoR Award 80NSSC22M0173. ACGC acknowledges the support from the GridUIS-2 experimental testbed. The latter was developed under the Universidad Industrial de Santander (SC3-UIS) High Performance and Scientific Computing Centre with support from UIS Vicerrectoría de Investigación y Extensión (VIE-UIS) and several UIS research groups. ACGC also acknowledge grant No. 202303059C entitled "Optimización de las Propiedades Termoeléctricas Mediante Tensión Biaxial en la Familia de Materiales $Bi_4O_4SeX_2$ ($X$ = Cl, Br, I) Desde Primeros Principios" supported by the LNS - BUAP.